\documentclass[traditabstract, letter]{aa} 
\usepackage{graphicx,natbib} 
\usepackage{txfonts} 
\bibpunct{(}{)}{;}{a}{}{,} 
 
\def\kms{km~s$^{-1}$}

\newcommand{\nc}{\newcommand}

\nc{\RAJ}[4]{$\alpha(J2000) = {#1}^{\rm h}{#2}^{\rm m}{#3}\fs{#4}$}
\nc{\DecJ}[4]{$\delta(J2000) = {#1}\degr {#2}\arcmin {#3}\farcs{#4}$}

\begin{document} 
 
\title{ALMA observations of the variable $^{12}$CO/$^{13}$CO ratio
  around the asymptotic giant branch star R Sculptoris \footnote{The
 total intensity image cubes are available
at the CDS via anonymous ftp to cdsarc.u-strasbg.fr (130.79.128.5)
or via http://cdsweb.u-strasbg.fr/cgi-bin/qcat?J/A+A/}}
 
\author{Vlemmings, W. H. T.\inst{1} 
\and Maercker, M.\inst{2,3} \and Lindqvist, M.\inst{1} \and Mohamed,
S.\inst{4} \and Olofsson,
H.\inst{1} \and Ramstedt, S.\inst{5} \and  Brunner, M.\inst{6} \and
Groenewegen, M. A. T.\inst{7} \and Kerschbaum, F.\inst{6} \and
 Wittkowski, M.\inst{3}}

\institute{Department of Earth and Space Sciences, Chalmers University of Technology, Onsala Space Observatory, SE-439~92 Onsala, Sweden
\email{wouter.vlemmings@chalmers.se} 
\and Argelander Institute f\"ur Astronomie, Universit\"at Bonn, 
Auf dem H\"ugel 71, 53121 Bonn, Germany\
\and European Southern Observatory, Karl Schwarzschild Str. 2,
Garching bei M\"unchen, Germany
\and South African Astronomical Observatory, PO Box 9, Observatory
7935, Cape Town, Western Cape, South Africa
\and Department of Physics and Astronomy, Division of Astronomy \&
Space Physics, Uppsala University, PO Box 516, 751~20 Uppsala, Sweden
\and University of Vienna, Department of Astrophysics,
T{\"u}rkenschanzstra\ss{}e 17, 1180, Wien, Austria
\and Koninklijke Sterrenwacht van Belgi{\"e}, Ringlaan 3, 1180, Brussels, Belgium
}
\date{received , accepted} 
\authorrunning{Vlemmings et al.} 
\titlerunning{The variable $^{12}$CO/$^{13}$CO ratio of R~Scl} 
 
\abstract {The $^{12}$CO/$^{13}$CO ratio is often used as a measure of
  the $^{12}$C/$^{13}$C ratio in the circumstellar environment,
  carrying important information about the stellar
  nucleosynthesis. External processes can change the $^{12}$CO and
  $^{13}$CO abundances, and spatially resolved studies of the
  $^{12}$CO/$^{13}$CO ratio are needed to quantify the effect of these
  processes on the
  globally determined values. Additionally, such studies provide
  important information on the conditions in the circumstellar
  environment. The detached-shell source R~Scl, displaying CO emission
  from recent mass loss, in a binary-induced spiral structure as
  well as in a clumpy shell produced during a thermal pulse, provides
  a unique laboratory for studying the differences in CO isotope
  abundances throughout its recent evolution. 
 We observed both the $^{12}$CO($J\,$=$\,$3$\,\to\,$2) and the $^{13}$CO($J\,$=$\,$3$\,\to\,$2) line
  using ALMA. We find significant variations in the
  $^{12}$CO/$^{13}$CO intensity ratios and consequently in the abundance
  ratios. The {\it average} CO isotope abundance ratio is at least a factor
  three lower in the shell ($\sim19$) than that in the present-day ($\lesssim300$ years) mass
  loss ($>60$). Additionally, variations in the ratio of more than an order of
  magnitude are found in the shell itself. We attribute these
  variations to the competition between selective dissociation and
  isotope fractionation in the shell, of which large parts
  cannot be warmer than $\sim35$~K. However, we also find that the
  $^{12}$CO/$^{13}$CO ratio in the present-day mass loss is significantly higher than the $^{12}$C/$^{13}$C ratio determined in
  the stellar photosphere from molecular tracers ($\sim19$). The origin of this discrepancy is
  still unclear, but we speculate that it is due to an embedded
  source of UV-radiation that is primarily photo-dissociating
  $^{13}$CO. This radiation source could be the hitherto hidden
  companion. Alternatively, the UV-radiation could originate from an
  active chromosphere of R~Scl itself. Our results indicate that
  caution should be taken when directly relating the
  $^{12}$CO/$^{13}$CO intensity and $^{12}$C/$^{13}$C abundance ratios
  for specific asymptotic giant branch stars, in particular binaries
  or stars that display signs of chromospheric stellar activity.}
 
\keywords{Stars: abundances,  Stars: AGB, Stars: carbon, Stars:
  circumstellar matter} 
 
\maketitle 
 
\section{Introduction} 

The study of different isotope ratios provides information on the
enrichment history of the interstellar medium. In particular the
$^{12}$C/$^{13}$C ratio carries an imprint of stellar evolution and
nucleosynthesis, as $^{12}$C is directly produced in the
triple-$\alpha$ process, while $^{13}$C is created from $^{12}$C as an
intermediate product of the carbon-nitrogen-oxygen (CNO) cycle. As the
importance of the CNO cycle increases for more massive stars, the
$^{12}$C/$^{13}$C ratio can be used to trace the past star formation
rate and stellar mass function \citep[e.g.][]{Prantzos96, Greaves97}.

Several studies have determined the photospheric
$^{12}$C/$^{13}$C ratio in evolved asymptotic giant branch (AGB) stars
\citep[e.g][]{Lambert86, Ohnaka96, Abia97}, by fitting
stellar atmosphere models to photospheric lines. There have also
been several papers in which the carbon isotope ratio is determined in
the circumstellar envelope itself, mainly by observations of the
$^{12}$CO/$^{13}$CO ratio
\cite[e.g.][]{Groenewegen96,Greaves97,Schoeier00,Milam09}. One of the
main conclusions from \citet{Schoeier00} is that the
$^{12}$CO/$^{13}$CO abundance ratio is a good proxy for determining
the $^{12}$C/$^{13}$C ratio provided accurate radiative transfer
modeling is performed. 

However, the aforementioned CO observations were all performed with
single-dish telescopes, which are insensitive to spatial variations in
the $^{12}$CO/$^{13}$CO intensity ratio. The effect of, for example,
an inhomogeneous circumstellar environment or (hidden) companion on the
$^{12}$CO/$^{13}$CO ratio throughout the circumstellar envelope is
thus poorly known. Previous submillimeter interferometer instruments
were not sensitive enough to map the often weak $^{13}$CO
emission at sufficient angular resolution. This has now changed with
the construction of ALMA, which for the first time provides sufficient
angular resolution and sensitivity to map the circumstellar $^{13}$CO
emission of AGB stars in detail.

Here we present $^{13}$CO($J\,$=$\,$3$\,\to\,$2) ALMA observations of the carbon-rich
AGB star R~Scl located at $\sim290$~pc \citep{Knapp03}. This star is
one of about a dozen detached-shell sources, where the shell is
thought to be created due to mass-loss modulation during a He-shell
flash (i.e., a thermal pulse) \citep{Olofsson90}.  The $^{12}$CO($J\,$=$\,$3$\,\to\,$2)
ALMA observations, taken together with the observations presented here,
revealed in addition to the shell an unexpected spiral connecting the
shell to the present-day ($\lesssim300$~yr) mass loss \citep[][hereafter
M+12]{Maercker12}. This spiral pattern is thought to be caused by
the motion of a previously unknown companion at $\sim60$~AU from R~Scl.

\begin{figure*}
\centering
\resizebox{0.67\hsize}{!}{\includegraphics{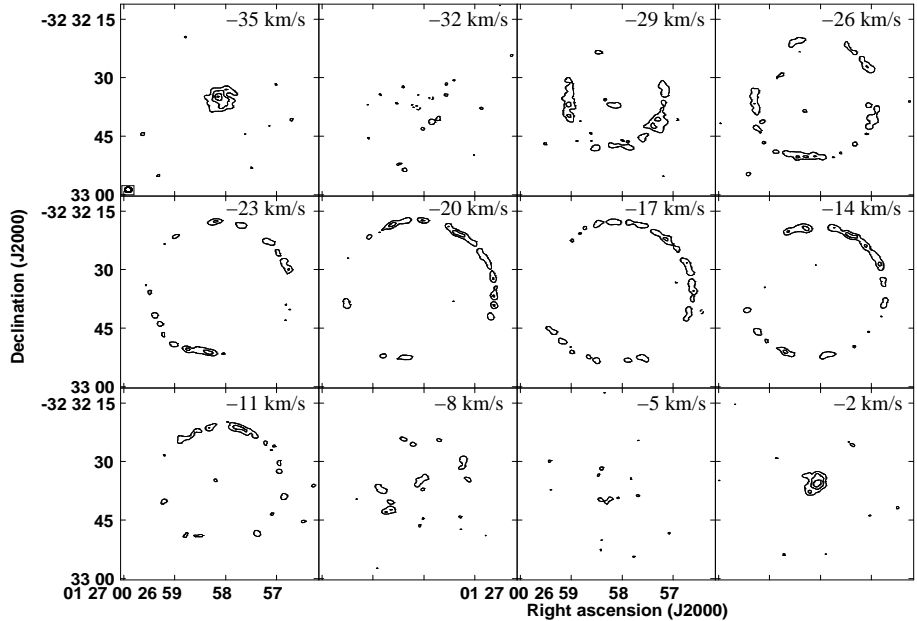}}
\caption{$^{13}$CO($J\,$=$\,$3$\,\to\,$2) intensity contours. The
  channels are averaged over three \kms. The panels are labeled
  with their $V_{\rm LSR}$ and the beam is indicated in the top left
  panel. The $^{13}$CO contour levels are drawn at $-4\sigma$ (dashed)
  and $4, 8, 12,$~and $16\sigma$, with
  $\sigma=15$~mJy~beam$^{-1}$.} \label{13CO}
\end{figure*}

\begin{figure*}
\centering
\resizebox{0.7\hsize}{!}{\includegraphics{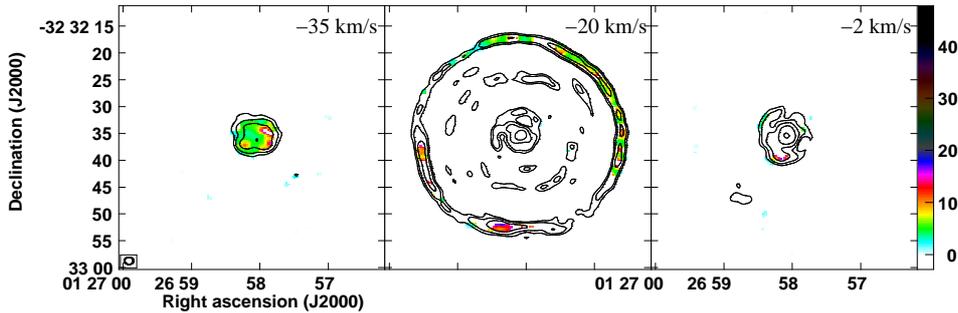}}
\caption{Intensity ratio, $I_{^{12}{\rm CO}} / I_{^{13}{\rm CO}}$, (color) and
$^{12}$CO($J\,$=$\,$3$\,\to\,$2) flux (contours) for three different velocity
channels (the full channel range is presented in Fig.~\ref{COratio2}). The
channels are averaged over three \kms~ and the panels are are labeled according to 
their $V_{\rm LSR}$. The ratio is presented where $^{13}$CO
emission is detected at higher than three $\sigma$ (where $\sigma=15$~
mJy~beam$^{-1}$) and the color scale runs from 0 to 42. The
$^{12}$CO contour levels are drawn at $6, 12, 24, 48$~and
$96\sigma$, with $\sigma=25$~mJy~beam$^{-1}$. Where no $^{13}$CO is
detected, the contours are equivalent to a $1\sigma$ lower limit to
the $^{12}$CO/$^{13}$CO intensity ratio of $10, 20, 40, 80,$~and
$160,$ respectively. The beam size is indicated in the left panel.} \label{COratio}
\end{figure*}

\section{ALMA observations and results}

The $^{12}$CO($J\,$=$\,$3$\,\to\,$2) and $^{13}$CO($J\,$=$\,$3$\,\to\,$2) emission lines of R Scl were
observed using ALMA Band 7 (275-373~GHz) on October 3--4 and 18--19
2011 during the ALMA cycle 0 observing program. The Band 7 data
contained four spectral windows of 1.875~GHz and 3840 channels each
that were tuned at 345.1~GHz, 343.3~GHz, 331.1~GHz, and 333.0~GHz. This
allowed us to simultaneously cover the $^{12}$CO($J\,$=$\,$3$\,\to\,$2) and
$^{13}$CO($J\,$=$\,$3$\,\to\,$2) at 345.795~GHz and 330.587~GHz, respectively.  The
results of the $^{12}$CO($J\,$=$\,$3$\,\to\,$2) observations were presented in M+12
and here we focus specifically on the $^{13}$CO($J\,$=$\,$3$\,\to\,$2) line. The
channel spacing of 0.488~MHz resulted after Hanning smoothing, in a
maximum spectral resolution of $0.44$~\kms. The data were taken using
the cycle 0 compact configuration of ALMA, with baselines ranging from
15~m to 200~m. Using natural weighting to optimize the sensitivity for
the weak $^{13}$CO emission, this resulted in a beam size of
$1.65"\times1.35"$ (at 330~GHz) with a position angle of 86$^\circ$. A
45-point mosaic, covering a region of $50"\times50"$, was made to
cover the entire shell of R~Scl. The center pointing of the mosaic was
taken to be the position of R~Scl at \RAJ{01}{26}{58}{094} and
\DecJ{-32}{32}{35}{454}. Each mosaic pointing was
observed for a total of 2.42~min and the total observing time was
approximately 3.8~hours. The data were reduced using the Common Astronomy Software Application
(CASA).  
After an initial correction for rapid atmospheric variations at
each antenna using water vapour radiometer data and correction for the
time and frequency dependence of the system temperatures, we improved
the antenna positions and performed a manual delay calibration.
 Bandpass calibration was performed on the
quasar 3C454.3 ($2.2$~Jy~beam$^{-1}$). The primary flux calibration
was made using Neptune and bootstrapping to the gain calibrator
J0137-245 ($0.49$~Jy~beam$^{-1}$). Based on the calibrator fluxes, the
absolute flux calibration has an uncertainty of $\sim10\%$.

%The data was reduced using the Common Astronomy Software Application
%(CASA). 
%After an initial correction for rapid atmospheric variations at
%each antenna using Water Vapour Radiometer data and correction for the
%time and frequency dependence of the system temperatures, we improved
%the antenna positions and performed a manual delay
%calibration. Subsequently, bandpass and gain calibration were
%performed, and the calibration solutions determined on J0137-245 were
%applied to R~Scl. 
Imaging was then done using the CASA clean
algorithm, smoothing the data to 3~\kms~to detect the weak
$^{13}$CO($J\,$=$\,$3$\,\to\,$2) emission. Similar smoothing was performed on the
$^{12}$CO($J\,$=$\,$3$\,\to\,$2) line for a direct comparison. The rms in the emission
line channels was $\sim15$~mJy~beam$^{-1}$ and
$\sim25$~mJy~beam$^{-1}$ for the $^{13}$CO($J\,$=$\,$3$\,\to\,$2) and $^{12}$CO($J\,$=$\,$3$\,\to\,$2)
lines, respectively. The increase in the rms noise for the
$^{12}$CO($J\,$=$\,$3$\,\to\,$2) line is due to the complex structure of the emission
(M+12). Because of the lack of short spacings, not all CO flux is
recovered. Emission at scales larger than $\sim15"$ is resolved
out. Based on single-dish observations of $^{12}$CO($J\,$=$\,$3$\,\to\,$2) and
$^{13}$CO($J\,$=$\,$3$\,\to\,$2) with the APEX telescope, we estimate $\sim25\%$ of the
emission is recovered for both transitions (see
  Appendix~\ref{sdish}). Our conclusions on the intensity ratio are
  thus not affected.

%\subsection{$^{13}$CO($J\,$=$\,$3$\,\to\,$2) molecular gas and intensity ratio maps} 

The $^{13}$CO($J\,$=$\,$3$\,\to\,$2) emission (hereafter denoted with only $^{13}$CO)
detected around R~Scl is shown in Fig.~\ref{13CO}. We find emission in
the detached shell, for which the $^{12}$CO($J\,$=$\,$3$\,\to\,$2) (hereafter
$^{12}$CO) maps are presented in M+12. The emission has a peak flux
in the shell of $\sim250$~mJy~beam$^{-1}$. However, the present-day
mass loss is only weakly detected (at
$4\sigma\approx60$~mJy~beam$^{-1}$) in $^{13}$CO in the $V_{\rm
  LSR}=-29, -11$, and $-8$~\kms~velocity channels.  Considering that the
detached molecular CO shell is relatively thin, the majority of the
flux will be resolved out toward the most red- and blue-shifted
peaks because this is where, in projection on the sky, the shell is
thickest. This can explain the relatively low level of detected
$^{13}$CO , as well as $^{12}$CO, in the shell at $V_{\rm LSR}=-32$
and $-5$~\kms.

In Fig.~\ref{COratio} and online Fig.~\ref{COratio2} we present the
measured intensity ratio, $I_{^{12}{\rm CO}} / I_{^{13}{\rm CO}}$,
around R~Scl together with the contours of the $^{12}$CO emission. While the
$^{13}$CO is not detected at all peaks of the $^{12}$CO shell
emission, the average $I_{^{12}{\rm CO}} / I_{^{13}{\rm CO}}$ is
significantly higher in the present-day mass loss than that in
the shell.

\section{Discussion} 
\label{discussion}

\subsection{The  $^{12}$CO/$^{13}$CO and $^{12}$C/$^{13}$C ratio}
 
If we now assume that the $^{12}$CO/$^{13}$CO-intensity ratio is
related to the $^{12}$C/$^{13}$C ratio, we can determine the latter
following \cite{Schoeier00}.  By comparing observations of the
$^{12}$CO/$^{13}$CO line intensity ratios with the
$^{12}$C/$^{13}$C ratios derived from \cite{Lambert86},
these authors determined a correlation of the form
${{I(^{12}{\rm CO})}\over{I(^{13}{\rm CO})}} =
(0.6\pm0.2)\times{{^{12}{\rm C}}\over{^{13}{\rm C}}}.$

Applying this relation to the average ratios found in the shell and
present-day mass loss, the carbon isotope ratio of R~Scl has increased
from $\sim12$ when the detached shell was formed to $\sim115$ during
the recent mass loss. However, these derived ratios do not take into
account optical depth effects, nor effects such as selective
photo-dissociation and chemical fractionation \citep[][]{Watson76}.
In particular, the carbon isotope ratio derived for the present-day mass
loss differs significantly from the value of $\sim19$
determined using molecular tracers in the stellar photosphere \citep{Lambert86}\footnote{A lower
  $^{12}$C/$^{13}$C ratio of $\sim9$ was determined in
  \cite{Ohnaka96}. \cite{Schoeier00}, however, found that their sample of
circumstellar $^{12}$CO/$^{13}$CO ratios are more consistent with the
work by \cite{Lambert86}.}.

To investigate more detailed radiative transfer effects
on the derived isotope ratios, we constructed radiative transfer
models as described in Appendix~\ref{radtrans}. The results of these are shown in
Fig.~\ref{ratio}. While our models can reproduce the double-peaked
emission from the detached shell using an {\it average}
$^{12}$CO/$^{13}$CO ratio of $19$ in the shell, the ALMA observations
require a ratio higher by at least a factor of three in the present-day
mass loss.

\subsection{Isotope ratio in the detached shell}

\begin{figure}
\centering
\resizebox{0.8\hsize}{!}{\includegraphics{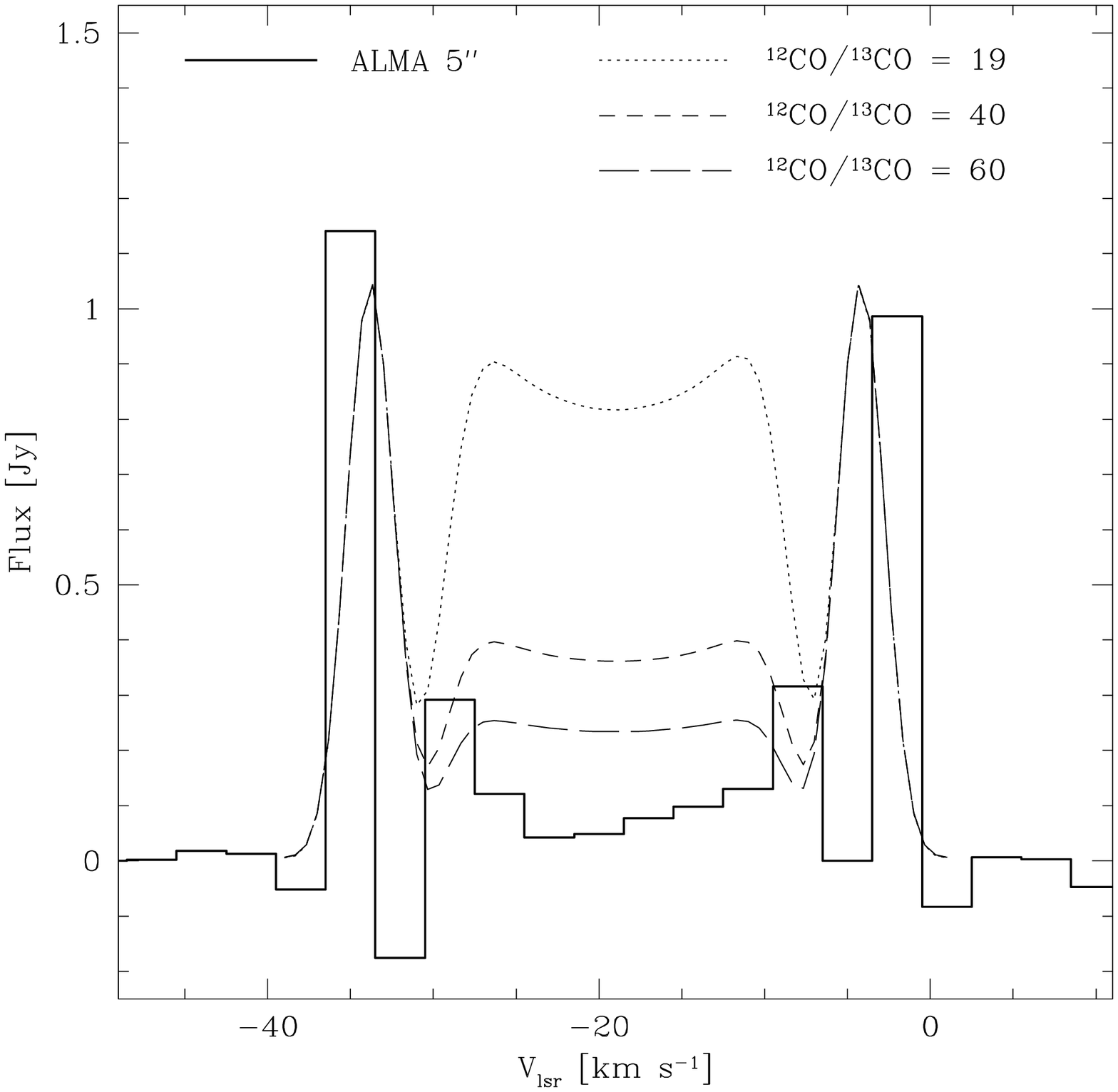}}
\caption{Histogram of the ALMA $^{13}$CO spectrum, convolved with
  a $5\arcsec$ beam to increase the signal-to-noise ratio, along the
  line-of-sight to R~Scl. The negative channels are due to significant
  missing flux near the front and back cap of the shell (see
  Appendix~\ref{sdish}), which does not affect the other channels with
  more compact emission. The dotted, short-dashed and long-dashed
  lines are three $^{13}$CO radiative transfer models (see
  Appendix~\ref{radtrans}) with different values for the
  $^{12}$CO/$^{13}$CO ratio only in the present-day mass loss. The
  ratio in the detached shell is kept at $19$.} \label{ratio}
\end{figure}

Our radiative transfer model can reproduce the {\it average}
$^{12}$CO/$^{13}$CO intensity ratio in the shell as shown in
Fig.~\ref{ratio} for an isotope ratio unchanged from the photospheric
value of $19$ found by \citet{Lambert86}. However, Fig.~\ref{COratio}
indicates that the intensity ratio ranges between as low as $\sim1.5$ to $>40$
where no $^{13}$CO is detected. This requires external processes, the
most likely of which are photo-dissociation and chemical
fractionation.  Changes in $^{12}$CO/$^{13}$CO ratio then occur
because, for photo-dissociation, self-shielding will
initially allow $^{12}$CO to be destroyed at a slower rate than
$^{13}$CO. However, at temperatures $\lesssim35$~K, chemical
fractionation would favor the creation of $^{13}$CO (and $^{12}$C$^+$)
from $^{12}$CO (and $^{13}$C$^+$) \citep[e.g][]{Watson76}.  While
\cite{Mamon88} showed that for regular AGB stars over a wide
mass-loss range, the difference in $^{12}$CO and $^{13}$CO abundance
is normally $\lesssim20\%$, \cite{Bergman94} showed that for the
detached-shell source S~Sct substantial fractionation is probably the
explanation for the strong detection of $^{13}$CO in that
source. Because of the high density in the shell of R~Scl, the model
in \citet{Bergman94} does not predict a significant effect from the
dissociation and fractionation.

However, it is clear from Figs.~\ref{13CO}~and~\ref{COratio} that the
shell of R~Scl is very clumpy. We find that the lowest $^{12}$CO/$^{13}$CO
intensity ratios are found to be anti-correlated with the strongest
$^{12}$CO peaks. This would be a natural consequence of the formation
of $^{13}$CO in the less dense and coldest regions ($<35$~K). The need
for the low temperature supports our estimate of $\sim50$~K for the
average shell temperature. We thus suggest that the CO isotope ratio
variation in the shell reflects the very inhomogeneous conditions in
the clumpy medium. This could further be confirmed by high-resolution
observations with ALMA of, for example, the neutral carbon
fine-structure lines, which should be correlated with the $^{13}$CO
emission peaks in the model described above.

\subsection{Isotope ratio in the present-day mass loss}

Our radiative transfer models furthermore indicate that the CO isotope
ratio in the present-day mass loss is $>60$. This is in contrast with
the photospheric $^{12}$C/$^{13}$C ratio of $\sim19$. It implies that,
contrary to the observations by \citet{Schoeier00}, the
$^{12}$CO/$^{13}$CO abundance ratio in the present-day mass loss of
R~Scl is not an accurate probe of the circumstellar $^{12}$C/$^{13}$C
ratio.  Alternatively, the photospheric
$^{12}$C/$^{13}$C ratio, derived from model-fitting lines of
  various molecular tracers, is more uncertain as expected or is not
representative for the carbon isotope ratio in the circumstellar
envelope. This last possibility is difficult to reconcile with the
circumstellar CO isotope observations that match the photospheric
values for many sources \citep[e.g.][]{Schoeier00, Milam09}.

Taken at face value, our results indicate a significant increase of
the $^{12}$C/$^{13}$C ratio after the detached shell was formed. Such
a carbon isotope ratio increase could potentially be a direct result of
the third dredge-up that occurred during the thermal pulse that
created the shell. During this dredge-up, convection in the hydrogen-burning shell reaches the He-burning shell, bringing $^{12}$C, created
by the triple-$\alpha$ process, to the surface. An initial isotope
abundance ratio of, in this case, $\sim19$, could then be increased to
$>40$. However, this scenario is unlikely for two main reasons: 1) it
would require the increased mass-loss and velocity that created the
shell to occur before the $^{12}$C is brought to the surface, which in
current models is impossible \citep[e.g.][]{Stancliffe04}, and 2) it
still does not explain the, in that case very recently decreased,
current photospheric carbon isotope ratio of $\sim19$. It is thus more
likely that the change in $^{12}$CO/$^{13}$CO ratio in the present-day
mass loss is due to external processes.

A speculative reason for the discrepancy between the CO and
C-isotope ratios is the presence of a strong UV-radiation source
within the circumstellar envelope. This source would then
predominantly dissociate $^{13}$CO, as $^{12}$CO would be
self-shielded. This strong source of UV-radiation could be the as yet
undetected companion of R~Scl responsible for creating the spiral
structure in the shell, or it could be the active chromosphere of
R~Scl itself. The strength of the UV-radiation would only need to be
similar to that of the interstellar radiation field and thus the
UV-radiation is not necessarily expected to penetrate the entire
circumstellar envelope and be directly detectable. \footnote{The GALEX
  catalog does however include a marginally significant UV-detection
  within $\sim10"$ of R~Scl with a near-uv flux of $\sim6~\mu$Jy.} Toward the outer
regions of the present-day mass-loss component, the $^{13}$CO would
then recombine, resulting in the observed {\it average} isotope ratio
of $\sim19$ in the shell. This hypothesis could be tested by observing
molecular species for which the dissociation is similar for both
isotopes, such as HCN or CS. Higher-resolution observations with
ALMA should also be able to determine regions in the near stellar
envelope that could be shielded from UV-radiation if this radiation
originates from the binary.

\section{Conclusions} 
\label{conclusions}

We found that the $^{12}$CO/$^{13}$CO ratio in the circumstellar
environment of the carbon AGB star R~Scl is significantly different in
the detached shell from the recent mass loss. The {\it
  average} ratio in the shell has decreased by at least a factor of
three compared with the ratio of $\gtrsim60$ determined for the material
within $<1000$~AU of the central star. The ALMA observations do reveal
strong variations of the $^{12}$CO/$^{13}$CO ratio in the shell
itself. These variations can very likely be attributed to the effect of
isotope fractionation dominating over the dissociation by interstellar
UV-radiation of $^{13}$CO in the strongly clumped shell
environment. However, the discrepancy between the ratio in the recent
mass loss and the observed atmospheric $^{12}$C/$^{13}$C ratio of
$\sim19$ \citep{Lambert86} is puzzling. One possible explanation could
be photo-dissociation of the less shielded $^{13}$CO by UV-radiation
from either chromospheric activity or from the thus far hidden
companion that is also responsible for the previously observed spiral
structure in the circumstellar envelope (M+12). In that case, one
would need to conclude that the relation between $^{12}$CO/$^{13}$CO
and $^{12}$C/$^{13}$C is not as straightforward as is often
assumed. Alternatively, the photospheric $^{12}$C/$^{13}$C ratio is
not a robust indicator for the $^{12}$C/$^{13}$C ratio of the
material expelled into the interstellar medium, or uncertainties in the
derived photosperic values are more significant than currently
thought. More observations with ALMA that can resolve the
circumstellar environment in the often much weaker $^{13}$CO lines and
other isotopes will be able to more firmly constrain the processes
involved, while improvements in models to derive the photospheric
isotope ratios are also being explored. Interestingly, if it is found
that the binary companion is the cause of the CO and C isotope ratio
discrepancy, such observations could also reveal hitherto hidden
companions to AGB stars.

%______________________________________________________________ 
 
\begin{acknowledgements} 
  This paper makes use of the following ALMA data:
  ADS/JAO.ALMA\#2011.0.00131.S . ALMA is a partnership of ESO
  (representing its member states), NSF (USA) and NINS (Japan),
  together with NRC (Canada) and NSC and ASIAA (Taiwan), in
  cooperation with the Republic of Chile. The Joint ALMA Observatory
  is operated by ESO, AUI/NRAO and NAOJ. WV was partly supported by
  the Deutsche Forschungsgemeinschaft (DFG; through the Emmy Noether
  Research grant VL 61/3-1) and by Marie Curie Career Integration Grant 321691. MB and FK acknowledge funding by the
  Austrian Science Fund FWF under project number P23586. We also
  thank the EU, Nordic and German ALMA regional centers for support
  during project planning and data analysis and John Black, Kjell
  Eriksson, Bengt Gustafsson and Richard Stancliffe for valuable
  discussion.
\end{acknowledgements} 
 
%______________________________________________________________ 

\begin{appendix}

\section{Comparison with single dish observations}
\label{sdish}

\begin{figure}
\centering
\resizebox{0.8\hsize}{!}{\includegraphics{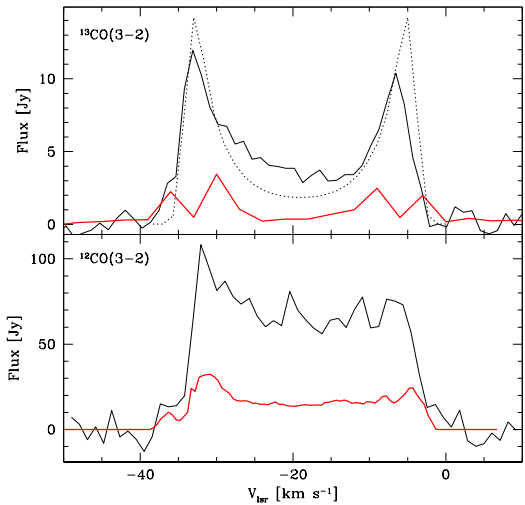}}
\caption{Total intensity of the $^{12}$CO and $^{13}$CO($J\,$=$\,$3$\,\to\,$2)
  transitions. The solid black lines are the spectra observed with the
  APEX telescope in a single pointing towards the star. The solid red
  line is the emission detected with ALMA in an area equivalent to the
  APEX beam. The $^{13}$CO ALMA spectrum has a velocity resolution of
  $3$~\kms, the $^{12}$CO ALMA spectrum has a velocity resolution of
  $0.5$~\kms, and the APEX spectra have a velocity resolution of
  $1.1$~\kms. The dotted line indicates the $^{13}$CO model used to
  fit the ALMA observations (see text) when observed with APEX
  resolution. The spectra indicate that the $^{12}$CO and $^{13}$CO
  emissions are equally resolved out in the ALMA
  observations.} \label{fluxloss}
\end{figure}

To determine if flux losses due to insufficient uv-coverage
of the ALMA observations could affect our conclusions, we compared the
interferometric observations with new spectra taken with the APEX
telescope. The $^{12}$CO and $^{13}$CO($\,$3$\,\to\,$2) observations were taken on
Jan 2, 2013 in a single pointing toward the star using the wobbler
with a $150"$ throw. The total bandwidth was 2.5~GHz, with 32768
channels, providing an initial resolution of $\sim0.1$~\kms, which was
later averaged to $\sim1.1$\kms. 

The resulting spectra are shown in Fig.~\ref{fluxloss} together with
the ALMA spectra taken over an area corresponding to the APEX beam at
the frequency of the CO($\,$3$\,\to\,$2) transition ($\sim19"$). From this figure
it is immediately obvious that a significant amount of flux is lost in
the ALMA observations. We only recover approximately $25\%$ of the
total flux, indicating that the circumstellar CO has significant
smooth emission over scales larger than $\sim15"$. Specifically in the
$^{13}$CO spectrum, most of the emission is lost toward the most red-
and blue-shifted peaks, where, as noted above, the caps of the shell
make up the largest area. A comparison of the observations with the
flux of the model used to fit the $^{13}$CO data, as shown in
Fig.~\ref{fluxloss} for a resolution of $19"$ and in Fig.~\ref{ratio} for
a resolution of $5"$, shows that most of the small-scale flux is
recovered. Considering that the present-day mass-loss extends out to
$<10"$, the lack of $^{13}$CO in the present-day mass-loss component
cannot be due to the flux loss in the interferometric
observations. Additionally, the single-dish comparison indicates that
the $^{12}$CO and $^{13}$CO are similarly resolved out, $25\%$ of the
  total flux is detected for both transitions, with
correspondingly little effect on the observed flux ratios. We thus find
that our conclusions based on the $^{12}$CO and $^{13}$CO intensity
ratios are robust.

\end{appendix}

\begin{appendix}

\section{Radiative transfer modeling}
\label{radtrans}

To more accurately estimate the $^{12}$CO/$^{13}$CO abundance ratio,
we performed radiative transfer calculations for the CO emission using
the code described in detail in \citet{Schoeier01}. Our model consists
of two components; a present-day mass loss and a detached shell.  The
density, velocity and temperature of the present-day mass loss (with
$\dot{M}=3\times10^{-7}$~M$_\odot$~yr$^{-1}$ derived from HCN
modeling) are taken to be the same as described in \citet{Schoeier05},
as are the stellar parameters. For the shell, we assumed a uniform gas
density with a shell width of $\sim 400$~AU and a total mass of
$2.5\times10^{-3}$~M$_\odot$. These parameters also match the models
of M+12, although there it was found that the increased mass-loss
during the thermal pulse did not decrease as rapidly as assumed here,
as evidenced by the spiral of CO gas connecting the shell with the
present-day mass loss. However, we do not include a description of the
spiral in our models because no $^{13}$CO was detected between the shell
and the more recent mass loss. We furthermore assumed an average
temperature of $50$~K for the shell. Because no observations of higher
rotational transitions of CO exist that can easily separate the
emission originating in the shell from that of the present-day mass
loss, the shell temperature is poorly constrained. A somewhat smaller
$^{13}$CO abundance and thus higher $^{12}$CO/$^{13}$CO ratio (by
$\sim20\%$) can reproduce the observed $^{13}$CO emission at higher
temperatures (up to $\sim70$~K). Finally, we only varied the $^{12}$CO
and $^{13}$CO abundance ratio in the present-day mass loss.

We do note that the derived $^{12}$CO/$^{13}$CO ratio for the present-day mass-loss is different from the value of $20$ derived from
modeling the single-dish CO($J\,$=2-1) observations
\citep{Schoeier00}. However, in that case it was assumed that the
ratio in the shell and present-day mass loss was the same, as with the
single-dish observations it is impossible to separate the two
components. The CO($J\,$=2-1) model also did not include the at that
time unknown spiral component. If we determine the average intensity
ratio over the entire envelope of R~Scl from the ALMA observations, we
find an intensity ratio of $10\pm3$, similar to the value of $12$
measured for the CO($J\,$=2-1) line \citep{Schoeier00}.   Furthemore, the new single-dish CO($J\,$=3-2) APEX observations
  yield an abundance ratio similar to that found in
  \citet{Schoeier00}, because, as was the case for the CO($J\,$=2-1)
  observations, the single-dish data cannot distinguish the detached
  shell and the present-day mass loss.

\end{appendix}

\begin{appendix}

\section{Channel maps}
\label{maps}

\begin{figure*}
\centering
\resizebox{\hsize}{!}{\includegraphics{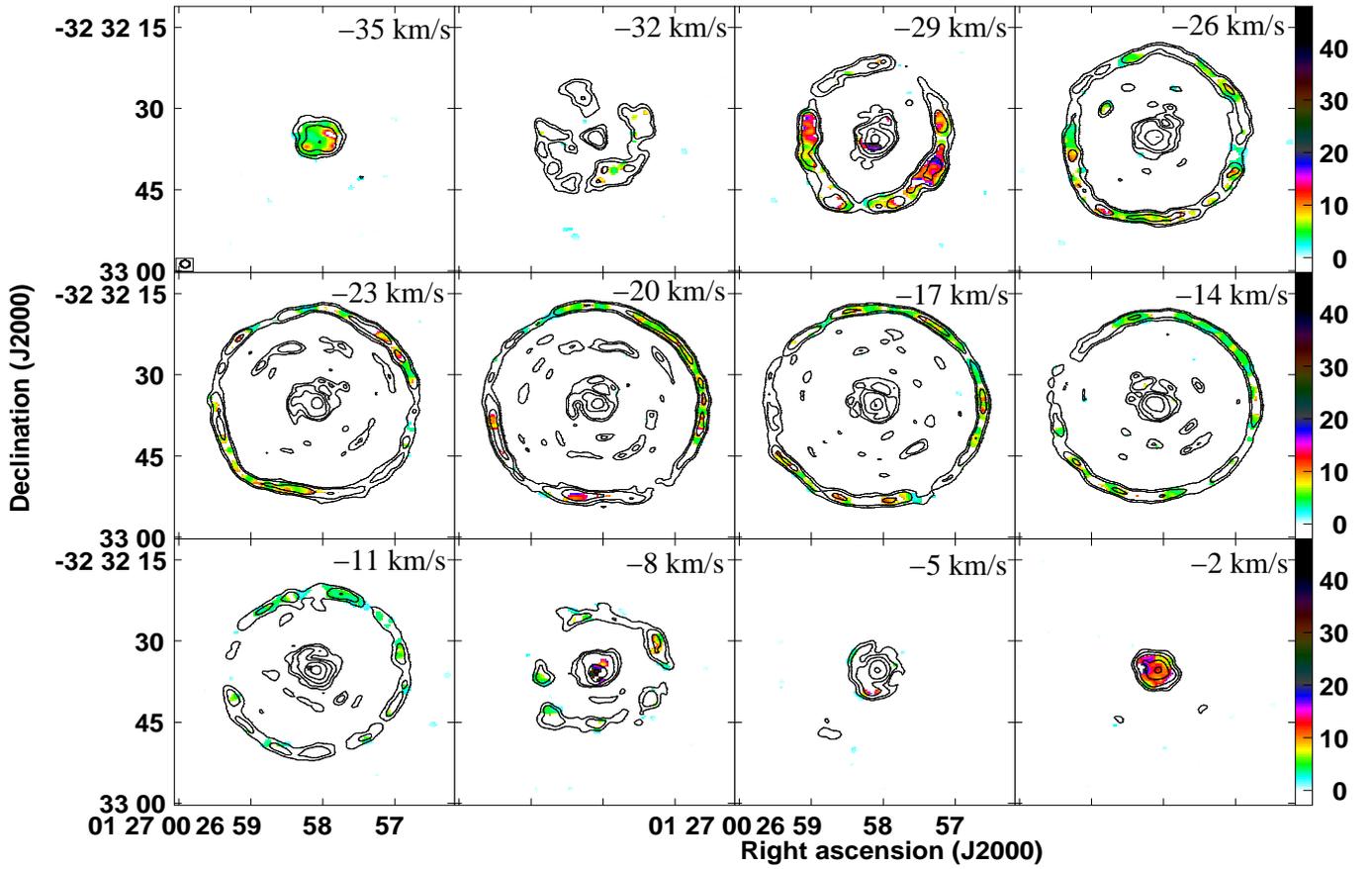}}
\caption{Intensity ratio, $I_{^{12}{\rm CO}} / I_{^{13}{\rm CO}}$, (color) and
$^{12}$CO($J\,$=$\,$3$\,\to\,$2) flux (contours) for the full velocity channel
range. The contour and color levels are as in Fig.~\ref{COratio}.} \label{COratio2}
\end{figure*}

\end{appendix}

\end{document}